\documentclass[preprint,aps,nofootinbib,showpacs,amsfonts,epsf]{revtex4}

\input{epsf.tex}

\newcommand{\E}{{\cal{E}}}

\renewcommand{\d}{{\rm d}}
\renewcommand{\a}{\alpha}

\newcommand{\be}{\begin{equation}}
\newcommand{\ee}{\end{equation}}
\newcommand{\bea}{\begin{eqnarray}}
\newcommand{\eea}{\end{eqnarray}}
\newcommand{\ba}{\begin{array}}
\newcommand{\ea}{\end{array}}
\def\J#1#2#3#4{{#1} {\bf #2}, #3 (#4)}
\def\PRD{Phys. Rev. D}
\def\PR{Phys. Rev.}
\def\PRL{Phys. Rev. Lett.}
\def\PTP{Prog. Theor. Phys.}

\def\JMP{J. Math. Phys.}

\def\TMP{Theor. Math. Phys.}

\def\CQG{Class. Quantum Grav.}

\def\PLA{Phys. Lett. A}

\begin{document}
\draft
\title{Exact solution for the simplest binary system\\ of Kerr black holes}

\author{V.~S.~Manko,$^*$ E.~D.~Rodchenko,$^\dagger$ E.~Ruiz$\,^\ddagger$ and B.~I.~Sadovnikov$\,^\dagger$}
\address{$^*$Departamento de F\'\i sica, Centro de Investigaci\'on y de
Estudios Avanzados del IPN, A.P. 14-740, 07000 M\'exico D.F.,
Mexico\\$^\dagger$Department of Quantum Statistics and Field Theory, Lomonosov Moscow State University, Moscow 119899, Russia\\$^\ddagger$Instituto Universitario de F\'{i}sica
Fundamental y Matem\'aticas, Universidad de Salamanca, 37008 Salamanca, Spain}

\begin{abstract}
The full metric describing two counter--rotating identical Kerr black holes separated by a massless strut is derived in the explicit analytical form. It contains three arbitrary parameters which are the Komar mass $M$, Komar angular momentum per unit mass $a$ of one of the black holes (the other has the same mass and equal but opposite angular momentum) and the coordinate distance $R$ between the centers of the horizons. In the limit of extreme black holes, the metric becomes a special member of the Kinnersly--Chitre five--parameter family of vacuum solutions generalizing the Tomimatsu--Sato $\delta=2$ spacetime, and we present the complete set of metrical fields defining this limit. \end{abstract}


\maketitle

\section{Introduction}

The binary systems of two aligned Kerr black holes are described by the well--known double--Kerr solution \cite{KNe}. It was proved \cite{MRu1} that two Kerr black holes with positive Komar masses cannot be in equilibrium due to balance of the gravitational and spin--spin forces without a supporting strut in between, the latter preventing the black holes from falling onto each other. In view of this it would be advantageous to identify a particular exact solution for the simplest binary system of Kerr black holes separated by a massless strut because such a solution would be free of pathological regions involving closed time--like curves and could probably find interesting applications in the black hole physics. It is worthwhile mentioning that in the paper \cite{MMR} an attempt was made to construct the simplest solution for a pair of stationary black holes by considering the corotating identical constituents. However, the constituents in the solution \cite{MMR} are separated by a special massive region which makes any accurate analysis of the thermodynamical properties of that specific binary system hardly possible. Fortunately, the axis condition is satisfied automatically in the case of another known exact solution constructed in \cite{BMa} and describing (in the absence of the charge parameter $q$) a binary system of counter--rotating identical Kerr black holes or hyperextreme objects. On the symmetry axis the Ernst complex potential $\E$ \cite{Ern} of the vacuum specialization of the Bret\'on--Manko solution has the form \be \E(\rho=0,z):=e(z)=\frac{(z-k'-m'-ia')(z+k'-m'+ia')} {(z-k'+m'-ia')(z+k'+m'+ia')}, \label{axis0} \ee the parameters $m'$, $a'$ and $k'$ being associated, respectively, with the mass, angular momentum and the separation distance. The whole set of the metric functions corresponding to the axis data (\ref{axis0}) which indeed define the simplest system of two Kerr black holes was given in \cite{BMa} in the explicit analytical form. At the same time, only the parameter $m'$ in (\ref{axis0}) is equal to the Komar mass \cite{Kom} of each constituent, whereas $a'$ does not represent exactly the angular momentum per unit mass (neither $k'$ is precisely the coordinate distance between the centers of the horizons). Obviously, the solution for the counter--rotating Kerr black holes will be physically more transparent if it is rewritten in terms of three physical parameters which are the mass $M$, angular momentum per unit mass $a$ and the relative coordinate distance $R$ {\it exactly}. Performing of such reparametrization and working out the limit of extreme counter--rotating Kerr black holes constitute the main two objectives of the present paper.

To accomplish the first goal, we shall make use of the expression for the irreducible mass $\sigma$ as function of $M$, $a$, $R$ obtained in \cite{Var} by solving the respective boundary Riemann--Hilbert problem, and also of the general theory of equatorially antisymmetric spacetimes developed in \cite{EMR}; this will permit us to reparametrize the axis data (\ref{axis0}) in terms of the parameters  $M$, $a$ and $R$ with the subsequent construction of the corresponding Ernst potential and metric functions by employing the general formulas of the paper \cite{MRu2}. The second objective will be achieved on the way of first identifying the asymptotically flat and equatorially antisymmetric member of the Kinnersley--Chitre family of vacuum solutions \cite{KCh} which provides the required limit of extreme Kerr black holes, and then the use of Yamazaki's formulas \cite{Yam} for working out the corresponding metric functions.

\section{The reparametrized non--extreme solution}

The solution derived in the paper \cite{BMa} involves the functions $r_i=\sqrt{\rho^2+(z-\a_i)^2}$ whose constants $\a_i$ determine the location of black holes on the symmetry axis (see Fig.~1). In our case of identical counter--rotating black holes we have $\a_1=-\a_4$, $\a_2=-\a_3$, and hence the constants $\a_i$ can be reparametrized as follows: \bea \a_1&=&{\textstyle\frac{1}{2}}R+\sigma, \quad \a_2={\textstyle\frac{1}{2}}R-\sigma, \quad \a_3=-{\textstyle\frac{1}{2}}R+\sigma, \quad \a_4=-{\textstyle\frac{1}{2}}R-\sigma, \quad \nonumber\\ R&=&{\textstyle\frac{1}{2}}(\a_1+\a_2-\a_3-\a_4), \quad \sigma={\textstyle\frac{1}{2}}(\a_1-\a_2)={\textstyle\frac{1}{2}}(\a_3-\a_4), \label{alfas} \eea where $R$ is the coordinate distance between the centers of the black hole horizons (the two rods in Fig.~1), and $\sigma$ is the so--called irreducible mass which is equal to the half length of each rod.

The crucial point now is to have the expression of $\sigma$ in terms of the physical quantities characterizing the binary system under consideration, which are the mass $M$, the angular momentum per unit mass $a$ of the lower black hole (the angular momentum of the upper black hole is $-Ma$), and the relative coordinate distance $R$. The desired expression for $\sigma$ was obtained in the paper \cite{Var} by Varzugin who studied the interaction force between the Kerr black holes via the boundary  Riemann--Hilbert problem posed on the symmetry axis. In the case of two counter--rotating identical black holes he was able to derive the following important relation: \be \sigma=\sqrt{M^2-a^2\mu}, \quad \mu:=\frac{R-2M}{R+2M}. \label{sigma} \ee

Using formula (\ref{sigma}), we can reparametrize the axis data (\ref{axis0}) in terms of the constants $M$, $a$ and $R$. Indeed, according to the general study of the equatorially symmetric/antisymmetric spacetimes \cite{EMR}, the axis data (\ref{axis0}) define the equatorially antisymmetric solution and hence can be written in the form \be e(z)=\frac{(z+\beta_1)(z+\beta_2)}{(z-\beta_1)(z-\beta_2)}, \label{axis1} \ee where the complex poles $\beta_1$ and $\beta_2$ satisfy the relation $\beta_1+\beta_2=-2M$. Recalling that the constants $\a_i$ are roots of the equation \be e(z)+\bar e(z)=0, \label{Sibga} \ee we find the explicit form of $\beta_1$ and $\beta_2$ from the equation \be e(z)+\bar e(z)= \frac{2(z-\a_1)(z-\a_2)(z-\a_3)(z-\a_4)} {(z-\beta_1)(z-\beta_2)(z-\bar\beta_1)(z-\bar\beta_2)} \label{ecu} \ee by equating the coefficients at the same powers of $z$ on the right--hand and left--hand sides of (\ref{ecu}). The result for $e(z)$ is the following simple expression: \be e(z)=\frac{z^2-2Mz-({\textstyle\frac12}R+M-ia)^2\mu} {z^2+2Mz-({\textstyle\frac12}R+M-ia)^2\mu}, \label{axis2} \ee which is the desired reparametrization of the data (\ref{axis0}) in physical parameters.

Once the axis data are established, the corresponding Ernst potential and the metric functions $f$, $\gamma$ and $\omega$ entering the stationary axisymmetric line element \be
\d s^2=f^{-1}[e^{2\gamma}(\d\rho^2+\d z^2)+\rho^2\d\varphi^2]-f(\d
t-\omega\d\varphi)^2 \label{Papa} \ee can be worked out with the aid of the general formulas obtained in the paper \cite{MRu2} for the analytically extended vacuum multi--soliton solution. Below we give the final expressions for $\E$, $f$, $\gamma$ and $\omega$ determining the simplest binary system of Kerr black holes: \bea \E&=&\frac{A-B}{A+B}, \quad f=\frac{A\bar A-B\bar B}{(A+B)(\bar A+\bar B)}, \quad e^{2\gamma}=\frac{A\bar A-B\bar B}{16R^4\sigma^4R_+R_-r_+r_-}, \nonumber\\ \omega&=&-\frac{2{\rm Im}[G(\bar A+\bar B)]}{A\bar A-B\bar B}, \nonumber\\ A&=&M^2[4\sigma^2(R_+R_-+r_+r_-)+R^2(R_+r_++R_-r_-)] \nonumber\\ &+&\{(R-2M)[R(\sigma^2-a^2)+2M^3]+4M^2a^2\mu\}(R_+r_-+R_-r_+) \nonumber\\ &-&2ia\sigma R(R-2M)(R_+r_--R_-r_+), \nonumber\\ B&=&2M\sigma R\{\sigma R(R_++R_-+r_++r_-)-[2M^2+ia(R-2M)](R_+-R_--r_++r_-)\}, \nonumber\\ G&=&-zB+M\sigma R\{2M[2\sigma(r_+r_--R_+R_-)+R(R_-r_--R_+r_+)] \nonumber\\ &+&(R+2\sigma)[R\sigma-2M^2-ia(R-2M)](R_+-r_-) \nonumber\\ &+&(R-2\sigma)[R\sigma+2M^2+ia(R-2M)](R_--r_+), \label{metric}  \eea where \be R_\pm=\sqrt{\rho^2+(z+{\textstyle\frac{1}{2}}R\pm\sigma)^2}, \quad r_\pm=\sqrt{\rho^2+(z-{\textstyle\frac{1}{2}}R\pm\sigma)^2}. \label{ri} \ee

Note that the antisymmetry of the metric (\ref{metric}) with respect to the equatorial ($z=0$) plane can be easily verified since under the transformation $z\to-z$ (i.e., $r_\pm\to R_\mp$, $R_\pm\to r_\mp$) the functions $f$ and $\gamma$ do not change, while $\omega(\rho,-z)=-\omega(\rho,z)$.

Turning now to the properties of black holes in our binary system, it should be mentioned that for each black hole the mass formula \cite{Sma} holds \be M=\frac{1}{4\pi}\kappa S+2\Omega J=\sigma+2\Omega J, \label{Sma} \ee where $\kappa$ is the surface gravity, $S$ the area of the horizon, $\Omega$ the angular velocity and $J$ the angular momentum. By construction, the mass $M$ enters the metric (\ref{metric}) as an arbitrary parameter, and so does $J$ through the parameter $a$ because the angular momentum of the lower black hole is $J=Ma$ (the upper black hole possesses the angular momentum $-J$ and the angular velocity $-\Omega$). The quantities $\kappa$, $S$ and $\Omega$ are defined on the horizon; for the calculation of $\kappa$ and $\Omega$ one can use the following simple formulas \cite{Tom}: \be \kappa=\sqrt{-\omega^{-2}e^{-2\gamma}}, \quad \Omega=\omega^{-1}, \label{kap} \ee where $\omega$ and $\gamma$ must be evaluated on the horizon. The expression for $\Omega$ was given by Varzugin in \cite{Var}, while $\kappa$ can be derived straightforwardly from (\ref{metric}): \be \Omega=\frac{a\mu}{2M(M+\sigma)}, \quad \kappa=\frac{R\sigma}{2M(M+\sigma)(R+2M)}. \label{Omega} \ee The expression for $S$ then follows from (\ref{Sma}) and (\ref{Omega}): \be S=\frac{4\pi\sigma}{\kappa}=8\pi M(M+\sigma)\Bigl(1+\frac{2M}{R}\Bigr). \label{S} \ee

The interaction force in the system of counter--rotating identical Kerr black holes was shown by Varzugin to be the same as in the case of two identical Schwarzschild black holes \cite{Var}: \be {\cal F}=\frac{M^2}{R^2-4M^2}. \label{F} \ee

The extreme limit of the solution (\ref{metric}) is defined by the equality $\sigma=0$, whence one finds \be R=\frac{2M(a^2+M^2)}{a^2-M^2}. \label{Rext} \ee This means, firstly, that for all $M>0$ the inequality $a^2>M^2$ holds for the extreme counter--rotating constituents, unlike in the case of a single Kerr black hole \cite{Ker} for which $a^2=M^2$; and, secondly, that, given $a^2>M^2$, there exists a distance (\ref{Rext}) at which the subextreme constituents become the extreme ones. At the same time, it should be pointed out that the limit $\sigma=0$ requires the application of l'Hospital rule in the formulas (\ref{metric}) and, therefore, needs a special treatment.\footnote{In a recent paper \cite{HRe} Herdeiro and Rebelo claim to have obtained a solution for two extremal counter--rotating Kerr black holes, but in reality their metric (3.8) has nothing to do with a genuine solution for such a binary system (however, for that case they correctly identify the inequality $M^2<a^2$ which is a trivial consequence of Varzugin's expression for $\sigma$).}

\section{The limit $\sigma=0$}

The simplest way of obtaining the solution for two extreme counter--rotating Kerr black holes consists in the use of the already known classical results after observing that in the limit $\sigma=0$ the metric (\ref{metric}) becomes a special member of the 5--parameter Kinnersley--Chitre family of vacuum spacetimes \cite{KCh} for which the whole set of metrical fields was given by Yamazaki \cite{Yam}. One only needs to find a concrete choice of the parameters in the Kinnersley--Chitre solution which leads to the axis data (\ref{axis2}) with $R$ defined by the formula (\ref{Rext}). The analysis of the behavior of the potential (4.30) of \cite{KCh} on the upper part of the symmetry axis ($y=1$) reveals that the required choice of the parameters in the original Kinnersley--Chitre formula (4.30) is the following: \be q=\beta=0, \quad p=1, \quad \a=\frac{2Ma}{a^2-M^2}, \quad \exp(-i\nu)=b+ic=\frac{a^2-M^2}{a^2+M^2}+i\frac{2Ma}{a^2+M^2}. \label{choice} \ee In addition, the prolate ellipsoidal coordinates $(x,y)$ must be related to the cylindrical coordinates $(\rho,z)$ via the formulas \be \rho=k\sqrt{(x^2-1)(1-y^2)}, \quad z=kxy, \quad k=\frac{M(a^2+M^2)}{a^2-M^2}. \label{xy} \ee

Below we give the final expressions for $\E$, $f$, $\gamma$ and $\omega$ describing the extreme counter--rotating Kerr black holes which were elaborated with the aid of Yamazaki's general formulas: \bea \E&=&\frac{A-B}{A+B}, \quad f=\frac{N}{D}, \quad e^{2\gamma}=\frac{N}{(1+\a^2)^2(x^2-y^2)^4}, \nonumber\\ \omega&=&4k\a y(x^2-1)(1-y^2)WN^{-1}, \nonumber\\ A&=&x^4-1+\a^2(x^2-y^2)^2-2i\a(x^2+y^2-2x^2y^2), \nonumber\\ B&=&2(b+ic)x[x^2-1-i\a(x^2-y^2)], \nonumber\\ N&=&[(x^2-1)^2+\a^2(x^2-y^2)^2]^2-16\a^2x^2y^2(x^2-1)(1-y^2), \nonumber\\ D&=&\{x^4-1+\a^2(x^2-y^2)^2+2x[b(x^2-1)+c\a(x^2-y^2)]\}^2 \nonumber\\ &+&4[\a(x^2+y^2-2x^2y^2)+cx(1-y^2)]^2, \nonumber\\ W&=&c\a(x^2-y^2)(3x^2+y^2)+b(3x^4+6x^2-1)+8x^3, \label{metric2} \eea where the constants $\a$, $b$, $c$ and $k$ have been defined in (\ref{choice}) and (\ref{xy}).

The metric (\ref{metric2}) describes two counter--rotating extreme Kerr black holes separated by the coordinate distance $2M(a^2+M^2)/(a^2-M^2)$. Mention that (\ref{metric2}) clearly shows that the metric function $\omega$ takes zero value both on the part $|z|>k$ of the symmetry axis (because $\omega(y=\pm1)=0$) and on the part $|z|<k$ of the axis (because $\omega(x=1)=0$).

\section{Conclusions}

In the present paper we have succeeded in bringing together various separate results on the simplest binary system of Kerr black holes for working out the most physical representation of the metric describing such system both in the subextreme case and in the limiting case of extremal black holes. It turns out that if the inequality $M^2>a^2$ holds, the black holes remain to be the subextreme constituents at any separation distance. On the other hand, if $M^2<a^2$, then there exists a critical distance defined by the formula (\ref{Rext}) before reaching which the black holes are subextreme objects; however, when this critical separation is achieved, the black holes become extremal, and at larger distances the constituents convert into the hyperextreme objects. The latter situation is also covered by the metric (\ref{metric}) because we have been working within the framework of Sibgatullin's method \cite{Sib} and analytically extended spacetimes \cite{RMM}.

Lastly, we remark that it would be also interesting to construct an exact solution for two corotating Kerr black holes separated by a massless strut, but this problem is technically more complicated than the one considered in the present paper and still remains a task for the future.

\section*{Acknowledgments}
This work was partially supported by Project 45946--F from CONACyT, Mexico, by Project FIS2006--05319 from MCyT, Spain, and by Project SA010C05 from Junta de Castilla y Le\'on, Spain.

\newpage

\begin{figure}[htb]
\centerline{\epsfysize=120mm\epsffile{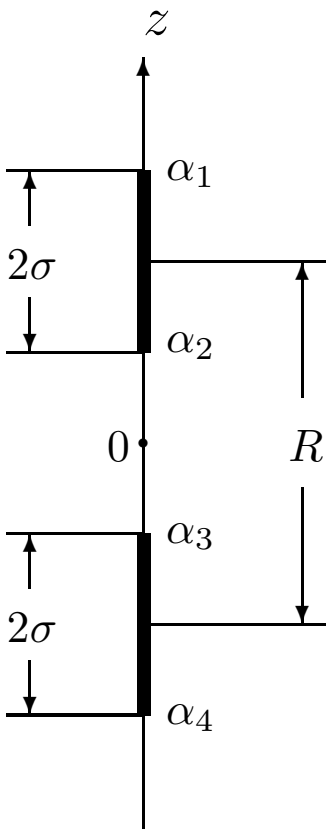}} \caption{The location of identical Kerr black holes on the symmetry axis: $\a_4=-\a_1$, $\a_3=-\a_2$.}
\end{figure}

\end{document}